\documentclass{jps-cp}
\usepackage{txfonts} %Please comment out this line unless the txfonts package is availabe in your LaTeX system.
\usepackage{color}%for preprints

\title{Effects of the Metallicity on Li and B Production in Supernova Neutrino Process}

\author{
Motohiko \textsc{Kusakabe}$^1$,
Myung-Ki \textsc{Cheoun}$^2$,
K. S. \textsc{Kim}$^3$,
Masa-aki \textsc{Hashimoto}$^4$,
Masaomi \textsc{Ono}$^5$,
Ken'ichi \textsc{Nomoto}$^6$,
Toshio \textsc{Suzuki}$^7$,
Toshitaka \textsc{Kajino}$^{1,8,9}$,
and
Grant J. \textsc{Mathews}$^{10}$
}

\inst{
$^1$School of Physics, and IRCBBC, Beihang University, Beijing 100083 China\\
$^2$Department of Physics and OMEG Institute, Soongsil University, Seoul 156-743, Korea\\
$^3$School of Liberal Arts and Science, Korea Aerospace University, Goyang 412-791, Korea\\
$^4$Kyushu University, Hakozaki, Fukuoka 812-8581, Japan\\
$^5$RIKEN, 2-1 Hirosawa, Wako-shi, Saitama 351-0198, Japan\\
$^6$Kavli IPMU (WPI), The University of Tokyo, Chiba 277-8583, Japan\\
$^7$Department of Physics, College of Humanities and Science, Nihon University, Tokyo 156-8550, Japan\\
$^8$National Astronomical Observatory of Japan, Tokyo 181-8588, Japan\\
$^9$Graduate School of Science, University of Tokyo, Tokyo 113-0033, Japan\\
$^{10}$Center for Astrophysics, Department of Physics, University of Notre Dame, IN 46556, U.S.A.}

\email{kusakabe@buaa.edu.cn}

\recdate{September 3, 2019}

\abst{The neutrino process ($\nu$-process) for the production of $^7$Li and $^{11}$B in core-collapse supernovae (SNe) is extensively investigated.
Initial abundances of $s$-nuclei and other physical conditions are derived from an updated calculation of the SN 1987A progenitor. The nuclear reaction network including neutrino reactions is constructed with the variable order Bader-Deuflhard integration method.
We find that yields of $^7$Li and $^{11}$B significantly depend on the stellar metallicity while they are independent of the weak $s$-process during the stellar evolution. When the metallicity is high, there are more neutron absorbers, i.e., $^{56}$Fe, $^{14}$N (from initial CNO nuclei), and $^{54}$Fe, and the neutron abundance is small during the $\nu$-process. Since $^7$Be is predominantly destroyed via $^7$Be($n$,$p$)$^7$Li, a change in the neutron abundance results in different $^7$Be yields. Then, the calculated yield ratio $^7$Li/$^{11}$B=0.93 for the solar metallicity is larger than that for the SN 1987A $^7$Li/$^{11}$B=0.80 by 16 \%
in the inverted mass hierarchy case. We analyze contributions of respective reactions as well as abundance evolution, and clarify the $\nu$-process of $^7$Li and $^{11}$B.}

\kword{astroparticle physics, neutrinos, nucleosynthesis, abundances, supernovae}

\begin{document}
\maketitle

\section{Introduction}

The neutrino spallation process of He and C in massive stars during Type II supernovae (SNe II) explosions contributes to the production of $^7$Li and $^{11}$B in the Galaxy \cite{1978Ap&SS..58..273D,1990ApJ...356..272W}. This process, as well as the $^3$He($\alpha$,$\gamma$)$^7$Be reaction \cite{1955ApJ...121..144C,1971ApJ...164..111C} in asymptotic giant branch stars \cite{2010MNRAS.402L..72V}, red giants \cite{1999ApJ...510..217S}, and novae \cite{1996ApJ...465L..27H}, and the production of LiBeB isotopes in cosmic ray nucleosynthesis via the spallation of CNO and $\alpha+\alpha$ fusion \cite{1970Natur.226..727R,1971A&A....15..337M}, is, therefore, important in Galactic chemical evolution of light elements.

SN explosions are energized by neutrinos emitted from the proto-neutron star (NS), and neutrino reactions with nuclei in the stellar interior produces a number of rare stable nuclei including $^{7}$Li, $^{11}$B, $^{138}$La, and $^{180}$Ta \cite{1978Ap&SS..58..273D,1990ApJ...356..272W,2005PhLB..606..258H} as well as short-lived nuclei $^{92}$Nb and $^{98}$Tc \cite{2012PhRvC..85f5807C,2013ApJ...779L...9H,Hayakawa2017}. Among them, production of the light elements Li and B occurs in the C- and He-rich layers inside of which neutrino flavor mixing can effectively occur due to the influence of the stellar electrons. Effects of neutrino oscillations have been clarified and Li and B yields have been calculated as functions of neutrino mixing angle $\sin^2 2 \theta_{13}$ for both cases of the normal and inverted mass hierarchies \cite{2006PhRvL..96i1101Y,2006ApJ...649..319Y}.

In this paper, we study effects of the initial nuclear abundances on the SN nucleosynthesis of $^7$Li and $^{11}$B. Production of those light nuclei in SNe is important because of the potential for constraining conditions of stellar evolution, and neutrino emission from NSs, and SN explosions. 

\section{Model}
Our model is described in Ref. \cite{kus18} in detail.
Presupernova density, temperature, and abundance profiles are taken from a new calculation using the method of \cite{2015PTEP.2015f3E01K} for the initial stellar mass $20 M_\odot$ with metallicity $Z=Z_\odot /4$ corresponding to SN1987A.
Hydrodynamical evolution of the SN is derived using the public code (blcode) (Christian Ott, Viktoriya Morozova, and Anthony L. Piro; https://stellarcollapse.org/snec). The explosion is triggered with a constant luminosity for 3 s with a total explosion energy of $10^{51}$ erg.
The nuclear reaction network code \cite{kus18} is based on the variable order Bader-Deuflhard method, and includes reaction rates from the JINA REACLIB database \cite{2010ApJS..189..240C} (Version 2.0).
The neutrino oscillations inside the star is solved with the MSW effect taken into account. Resulting flavor change probabilities are then utilized in deriving rates of reactions between neutrinos and nuclei.
Neutrino temperatures are $T_{\nu_e} =3.2$, $T_{\bar{{\nu_e}}} =5.0$, and $T_{\nu_x} =6.0$ MeV (for $\nu_x$ =$\nu_\mu$, $\nu_\tau$, $\bar{\nu_\mu}$, and $\bar{\nu_\tau}$) \cite{2004ApJ...600..204Y}.
In this calculation, we take into account dependence of the neutral current cross sections for the spallation of $^4$He and $^{12}$C on neutrinos and antineutrinos, which has been neglected in previous studies \cite{2008ApJ...686..448Y}, based on the models WBP and SFO \cite{2008ApJ...686..448Y}.

\section{Results}

Important reactions in five typical layers of SNe II have been shown, and details on effects of neutrino oscillations and the initial nuclear abundances have been investigated \cite{kus18}. Here we show how $^7$Li and $^{11}$B production in the SN $\nu$-process is affected by the stellar metallicity.

\begin{figure}[tbh]
  \begin{center}
    \includegraphics[width=0.45\textwidth]{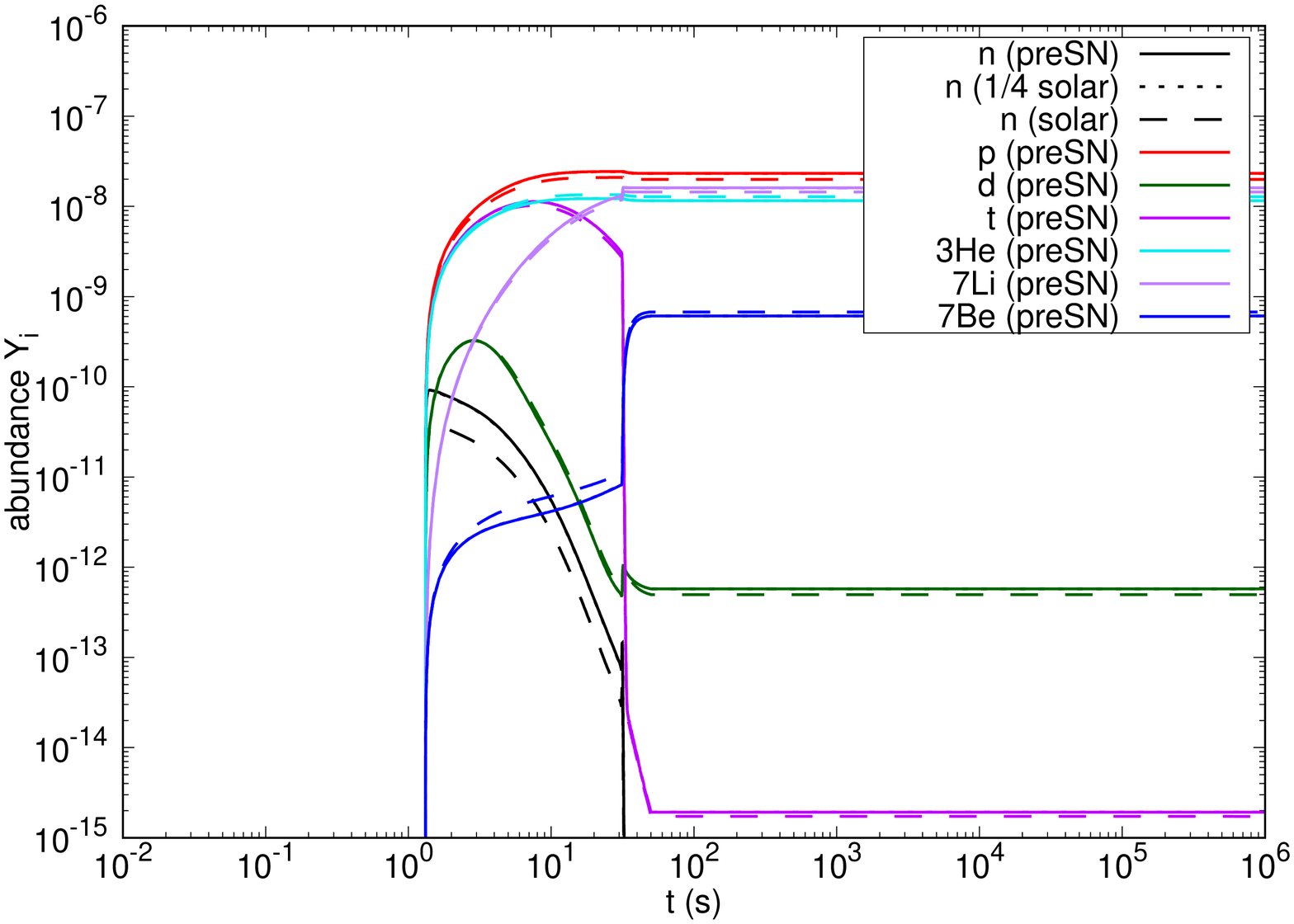}
    \includegraphics[width=0.45\textwidth]{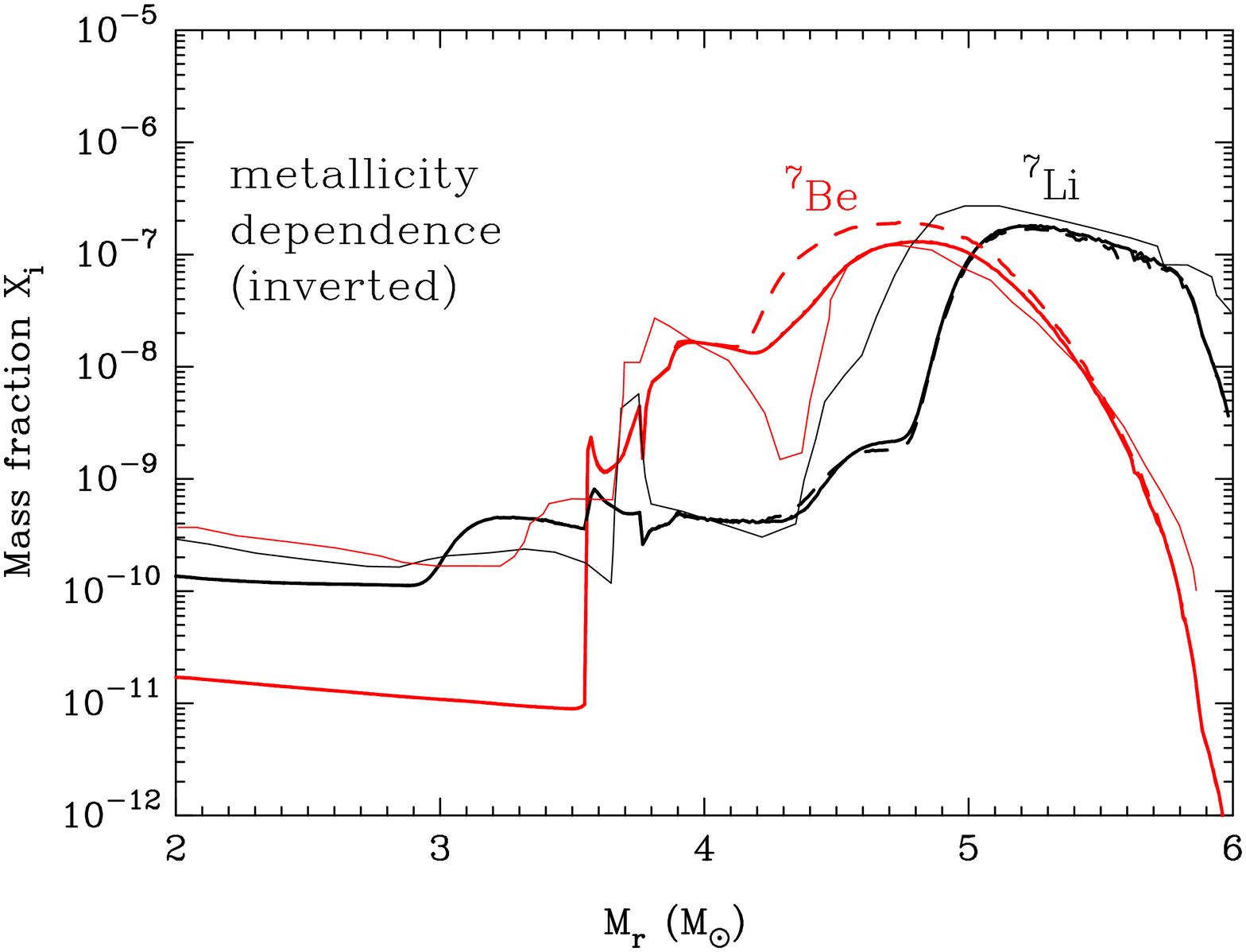}
  \end{center}
  \caption{(Left panel) Nuclear abundances as a function of time at $M_r=5 M_\odot$. Solid, dotted, and dashed lines correspond to cases of initial abundances for heavy elements from the standard $s$-process during stellar evolution of $Z=Z_\odot/4$ (case 2), the solar abundances divided by 4 (case 4), and solar abundances (case 5), respectively.
(Right panel) Final mass fractions of $^7$Li and $^7$Be versus Lagrangian mass coordinate. Reprinted from Ref. \cite{kus18}. Thick and thin lines correspond to the current results \cite{kus18} and the previous results \cite{2006ApJ...649..319Y}, respectively.
      The case numbers refer to those in Ref. \cite{kus18}. Dotted lines are overlapping with solid lines.}
\label{f1}
\end{figure}

Figure \ref{f1} (left panel) shows nuclear abundances as a function of time at $M_r =5 M_\odot$ for the inverted neutrino mass hierarchy case. Abundances of $n$, $p$, $d$, $t$, $^3$He, $^7$Li, and $^7$Be are plotted.
First, the neutrino emitted at the NS surface arrives in this region at $t \sim 1$ s, and the neutrino spallation of $^4$He operates. Nuclei produced via the spallation starts nucleosynthesis. Before the shock arrives at $t \sim 30$ s, the abundances of $^3$He and $^7$Be are largest and those of protons and neutrons are smallest in case 5.
At the shock arrival, the gas is heated and the $^7$Be abundance is slightly smaller while the $^7$Li abundance is slightly larger in case 2.
The neutron abundance is determined by abundances of neutron absorbers, i.e., the metallicity. In a metal-rich environment, the neutron abundance during the SN is small as seen in the left panel.
We note that although  $^{11}$B and $^{11}$C are also produced via the $\nu$+$^{12}$C reaction, no difference is observed in their abundances between the three cases.

Figure \ref{f1} (right panel) shows final mass fractions of $^7$Li, and $^7$Be as a function of the Lagrangian mass coordinate (thick lines) \cite{kus18}. The inverted mass hierarchy has been assumed for the three models. 
There is no significant difference in abundances of $^7$Li and $^{11}$B (see fig. 14 in Ref. \cite{kus18}) in the He-rich layers ($M_r \gtrsim 3.9 M_\odot$). However, the difference in the $^7$Be abundance is rather large. At the peak of the yield, the $^7$Be abundance is the largest in case 5.

\begin{figure}[tbh]
  \begin{center}
    \includegraphics[width=0.45\textwidth]{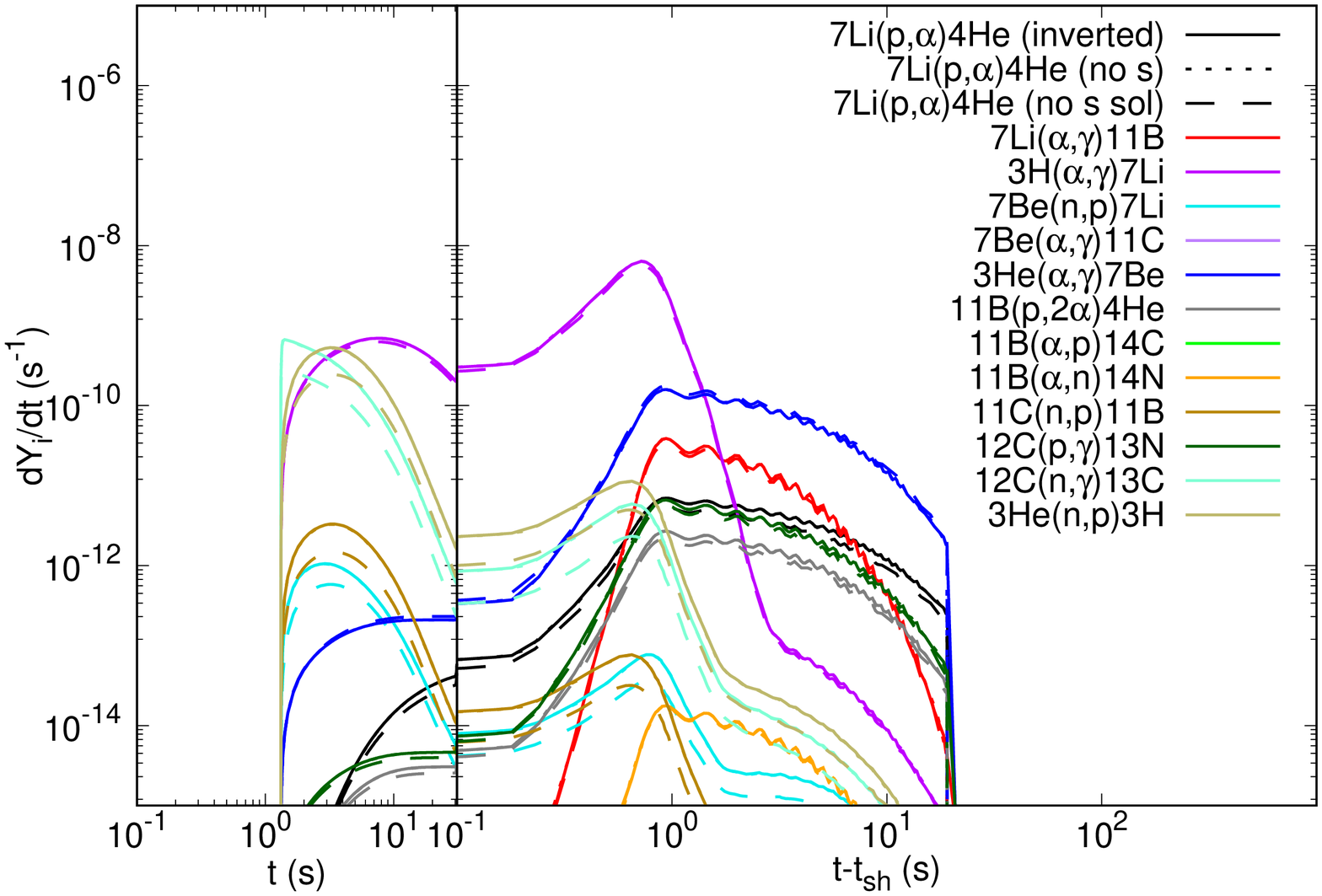}
    \includegraphics[width=0.45\textwidth]{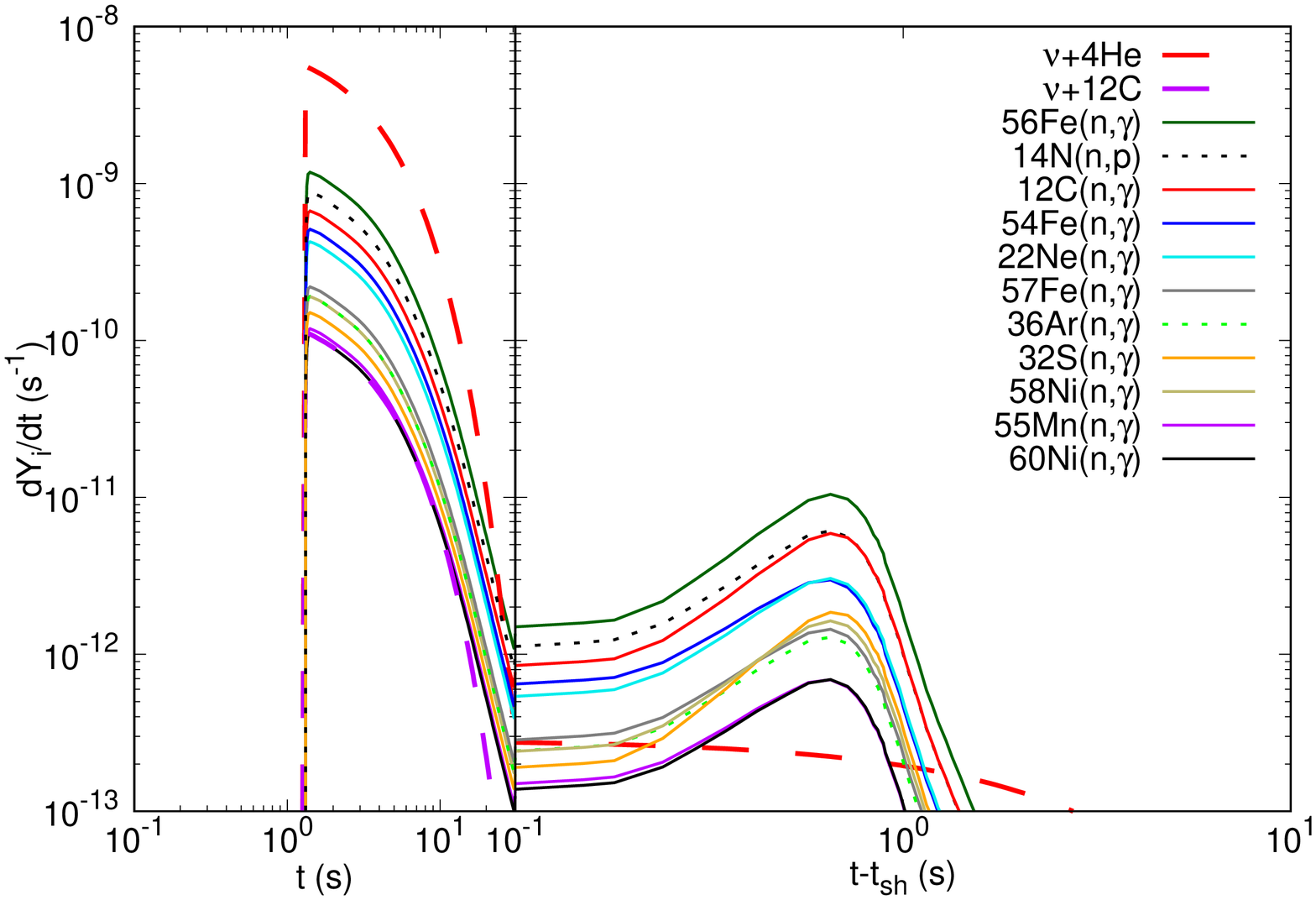}
  \end{center}
  \caption{Rates of abundance change $dY_i/dt$ via respective reactions versus time at $M_r=5 M_\odot$:
    (a) light nuclei in the initial states of two-body nuclear reactions in the same cases as in Fig. \ref{f1}; (b) main production and destruction reactions of neutrons only in case 2.
     The vertical line corresponds to the time of shock passage at $t= t_{\rm sh}$.} 
\label{f2}
\end{figure}

Figure \ref{f2} (left panel) shows the rates of abundance changes $dY_i/dt$ at $M_r =5 M_\odot$  for the inverted hierarchy case. Solid and dashed lines correspond to results for cases 2 and 5, respectively. Lines for case 4 are almost the same as those for case 2 in the whole stellar region.
After the neutrino arrival at $t \sim 1$ s, nucleosynthesis is initiated by the neutrino spallation of $^4$He. However, because of a relatively low temperature, charged particle reactions of heavy nuclei do not operate effectively. At the shock passage at $t \sim 30$ s, the temperature is increased, and rates of nuclear reactions including $^3$H($\alpha$,$\gamma$)$^7$Li, $^3$He($\alpha$,$\gamma$)$^7$Be, and $^7$Li($\alpha$,$\gamma$)$^{11}$B are sufficiently increased. Because of the stronger neutron absorption in metal-rich conditions, a lower neutron abundance is realized, and all rates for neutron captures $^3$He($n$,$p$)$^3$H, $^7$Be($n$,$p$)$^7$Li, and $^{12}$C($n$,$\gamma$)$^{13}$C are smallest in case 5.
Initial stellar metallicity never changes abundances of $^4$He and $^{12}$C in this outer region. Therefore, the abundance change rates of neutrino spallation reactions do not depend on the metallicity.

Figure \ref{f2} (right panel) shows production (dashed lines) and destruction rates (solid and dotted lines) for neutron, i.e., $|dY_n/dt|$ via main reactions, as a function of time at $M_r=5 M_\odot$. Once the neutrino arrives at this layer, neutrons are produced via the $\nu+^4$He spallation reaction, and various nuclei from carbon to iron-peak elements captures the neutrons via $(n,\gamma)$ reactions plus $^{14}$N$(n,p)$ and $^3$He($n$,$p$) (Fig. \ref{f2}a). The total rates of destruction via $n$-capture and production via the neutrino spallation are completely balanced until $t \sim 10$ s. However, because of the decreasing neutrino luminosity, the destruction rate eventually dominates for $t \gtrsim 10$ s. At the shock passage, the density and temperature are temporal lily increased and the neutron capture rates have peaks at $t-t_{\rm sh} \lesssim 10$ s.

  In this He-rich layer of massive stars, the neutrino spallation of $^4$He induces neutron capture reactions including $^7$Be($n$,$p$)$^7$Li. Many abundant nuclei in this layer, i.e., $^{56}$Fe, $^{14}$N, $^{54}$Fe, and so on, originate from the initial stellar metallicity, and contribute as neutron poisons during the synthesis of $^7$Li and $^7$Be. Thus, the $^7$Be abundance from the SN $\nu$-process depends on the stellar metallicity (via the abundance of neutron poison) and the neutrino mass hierarchy (via the production rates of $\nu$-spallation yields \cite{kus18})
.

%\subsection{Metallicity dependence}

%\begin{figure}[tbh]
%  \begin{center}
%    \includegraphics[width=0.40\textwidth]{li7_metal.eps}
%    \includegraphics[width=0.40\textwidth]{b11_metal.eps}
%  \end{center}
%  \caption{Mass fractions of $^7$Li and $^7$Be (left panel) and $^{11}$B and $^{11}$C (right panel) versus Lagrangian mass coordinate. Thick and thin lines correspond to the current results \cite{kus18} and the previous results \cite{2006ApJ...649..319Y}, respectively. The solid and dashed lines correspond to case 2 (the presupernova $s$-abundances for $Z=Z_\odot /4$) and case 5 (solar abundances), respectively. Reprinted from Ref. \cite{kus18}.}
%\label{f3}
%\end{figure}

The $\nu$-process in SNe provides an important contribution to the Galactic chemical evolution of $^7$Li and $^{11}$B \cite{pra12}.
Therefore, in the Galactic chemical evolution, the metallicity effect on the $\nu$-process must be taken into account. Also, yields of $^7$Li and $^{11}$B in SNe with near-solar metallicities, not that of SN 1987A progenitor, should be utilized when the calculated $^7$Li/$^{11}$B ratio is compared with meteoritic data for presolar grains \cite{mat12}.

\section{Parameter dependence}

Neutrino energy spectra and their time evolution as well as the explosion energy affect the yields of $^7$Li and $^{11}$B \cite{2004ApJ...600..204Y,2005PhRvL..94w1101Y,2006PhRvL..96i1101Y,2006ApJ...649..319Y,2008ApJ...686..448Y}. The $^7$Li and $^{11}$B abundances are also affected by collective neutrino oscillation effects in the inner region of SNe \cite{Ko2018}. Recent results from SN numerical simulations indicate rather similar temperatures among different flavors of neutrinos, and time evolution of flavor-dependent neutrino luminosities. These additional effects drastically change the yields of $^7$Li and $^{11}$B as well as other heavy rare nuclei such as $^{138}$La and $^{180}$Ta and short-lived nuclei $^{92}$Nb and $^{98}$Tc (to appear in an updated version of Ref. \cite{Ko2018}).


\begin{thebibliography}{99}
\bibitem{1978Ap&SS..58..273D} G.~V. Domogatskii, R.~A.Eramzhian, and D.~K. Nadezhin, Astrophys. Space Sci. {\bf 58}, 273 (1978).
\bibitem{1990ApJ...356..272W} S.~E. Woosley, D.~H. Hartmann, R.~D. Hoffman, and W.~C. Haxton, Astrophys. J. {\bf 356}, 272 (1990).
\bibitem{1955ApJ...121..144C} A.~G.~W. Cameron, Astrophys. J. {\bf 121}, 144 (1955).
\bibitem{1971ApJ...164..111C} A.~G.~W. Cameron, and W.~A. Fowler, Astrophys. J. {\bf 164}, 111 (1971).
\bibitem{2010MNRAS.402L..72V} P. Ventura, and F. D'Antona, Mon. Not. R. Astron. Soc. {\bf 402}, L72 (2010).
\bibitem{1999ApJ...510..217S} I.-J. Sackmann, and A.~I. Boothroyd, Astrophys. J. {\bf 510}, 217  (1999).
\bibitem{1996ApJ...465L..27H} M. Hernanz, J. Jose, A. Coc, and J. Isern, Astrophys. J. Lett. {\bf 465}, L27 (1996).
\bibitem{1970Natur.226..727R}  H. Reeves, Nature {\bf 226}, 727 (1970).
\bibitem{1971A&A....15..337M}  M. Meneguzzi, J. Audouze, and H. Reeves, Astron. Astrophys. {\bf 15}, 337  (1971).
\bibitem{2005PhLB..606..258H}  A. Heger, E. Kolbe, W.~C. Haxton, {\it et al.}, Phys. Lett. B, {\bf 606}, 258 (2005).
\bibitem{2012PhRvC..85f5807C}  M.-K. Cheoun, E. Ha, T. Hayakawa, {\it et al.}, Phys. Rev. C {\bf 85}, 065807  (2012).
\bibitem{2013ApJ...779L...9H}  T. Hayakawa, K. Nakamura, T. Kajino, {\it et al.}, Astrophys. J. Lett. {\bf 779}, L9  (2013).
\bibitem{Hayakawa2017}  T. Hayakawa, H. Ko, M.~K. Cheoun, {\it et al.}, Phys. Rev. Lett. {\bf 121}, 102701 (2018).
\bibitem{2006ApJ...649..319Y} T. Yoshida, T. Kajino, H. Yokomakura, {\it et al.}, Astrophys. J. {\bf 649}, 319  (2006).
\bibitem{2006PhRvL..96i1101Y} T. Yoshida, T. Kajino, H. Yokomakura, {\it et al.}, Phys. Rev. Lett. {\bf 96}, 091101 (2006).
\bibitem{kus18}  M.~Kusakabe, M.~K. Cheoun, K. S. Kim, {\it et al.},
  %``Supernova Neutrino Process of Li and B Revisited,''
  Astrophys.\ J.\  {\bf 872}, 164 (2019).
%  doi:10.3847/1538-4357/aafc35
%  [arXiv:1901.01715 [astro-ph.HE]].
  %%CITATION = doi:10.3847/1538-4357/aafc35;%%
  %2 citations counted in INSPIRE as of 01 Sep 2019

\bibitem{2015PTEP.2015f3E01K} Y. Kikuchi, M.-a. Hashimoto, M. Ono, and R. Fukuda, Prog. Theor. Exp. Phys. {\bf 063E01}  (2015).
\bibitem{2010ApJS..189..240C}  R.~H. Cyburt, A.~M. Amthor, R. Ferguson, {\it et al.}, Astrophys. J. Suppl. {\bf 189}, 240  (2010).
\bibitem{2004ApJ...600..204Y} T. Yoshida, M. Terasawa, T. Kajino, and K. Sumiyoshi, Astrophys. J. {\bf 600}, 204 (2004).
\bibitem{2008ApJ...686..448Y} T. Yoshida, T. Suzuki, S. Chiba, {\it et al.}, Astrophys. J. {\bf 686}, 448 (2008).
\bibitem{pra12} N.~Prantzos, Astron.\ Astrophy.\ {\bf 542}, A67 (2012).
\bibitem{mat12} G.~J.~Mathews, T.~Kajino, W.~Aoki, {\it et al.}, Phys. Rev. D {\bf 85}, 105023 (2012).

\bibitem{2005PhRvL..94w1101Y}  T. Yoshida, T. Kajino, and D.~H. Hartmann, Phys. Rev. Lett. {\bf 94}, 231101 (2005).
\bibitem{Ko2018} 
  H.~Ko, M.-K. Cheoun, E. Ha, {\it et al.},
  %``Neutrino self-interaction and MSW effects on the supernova neutrino-process,''
  arXiv:1903.02086 [astro-ph.HE].
  %%CITATION = ARXIV:1903.02086;%%
  %1 citations counted in INSPIRE as of 01 Sep 2019
\end{thebibliography}
\end{document}